\documentclass[twocolumn,showpacs,preprintnumbers]{revtex4}%
\usepackage[utf8]{inputenc}
\usepackage{amssymb}
\usepackage{amsmath}
\usepackage{float}
\usepackage{units}
\usepackage{graphicx}
\usepackage{epstopdf}
\usepackage{dcolumn}
\usepackage{comment}
\usepackage{bm}
\usepackage{xcolor}
\usepackage{soul}
\usepackage{hyperref}
\usepackage{amsfonts}%
\begin{document}
\title{Tunable degree of polarization in a Figure-8 fiber laser}
\author{Banoj Kumar Nayak}
\affiliation{Andrew and Erna Viterbi Department of Electrical Engineering, Technion, Haifa
32000 Israel}
\author{Cijy Mathai}
\affiliation{Andrew and Erna Viterbi Department of Electrical Engineering, Technion, Haifa
32000 Israel}
\author{Dmitry Panna}
\affiliation{Andrew and Erna Viterbi Department of Electrical Engineering, Technion, Haifa
32000 Israel}
\author{Eyal Buks}
\email[corresponding author: ]{eyal@ee.technion.ac.il}
\affiliation{Andrew and Erna Viterbi Department of Electrical Engineering, Technion, Haifa
32000 Israel}
\date{\today }

\begin{abstract}
We experimentally study a fiber loop laser in the Figure-8 configuration, and
explore the dependency of the degree of polarization on controlled parameters.
To account for the experimental observations, a mapping is derived to evaluate
the polarization time evolution. Nonlinearity induced by Kerr effect and gain
saturation gives rise to rich dynamics. We find that degree of polarization
can be increased by tuning the system into a region where the mapping has a
locally stable fixed point.

\end{abstract}
\maketitle

\section{Introduction}

Scrambling is commonly employed to lower the degree of polarization (DOP) of
optical sources, when effects such as polarization hole-burning and
polarization mode dispersion are unwanted
\cite{Cheng_1031,Safari_2865,Barrera_20268}. On the other hand,
high DOP is desired for some other applications, including the detection of
magneto-optic \cite{Duggan_1152, Ali_035001} and optomechanical effects
\cite{He_e1600485}.

In this study we explore the DOP of a fiber loop laser in the Figure-8
configuration. Mode locking (ML) in Figure-8 lasers (F8L) has been studied in
\cite{Duling_539,Lin_1121,Liao_14705,Kuse_3095,Hansel_331}. Stability of a passively mode-locked F8L was studied in \cite{Salhi_033828}.
Various types of vector solitary waves have been demonstrated in optical
fibers \cite{Menyuk_614,Menyuk_392,Afanasjev_270,Christodoulides_53}, and
mode-locked fiber lasers \cite{Akhmediev_852,Cundiff_3988}. Polarization
evolution of vector solitary waves have been investigated in optical fibers
\cite{Silberberg_246,Barad_3290,Cundiff_3988} and fiber lasers
\cite{Sergeyev_e131,Collings_354,Zhang_2317,Zhao_10053,Zhang_052302,Tang_153904,Mou_3831,Wang_413}%
.

Polarization evolution in optical fibers with spatially varying birefringence
has been explored in \cite{Menyuk_1288}. Polarization evolution of solitary
waves in fiber lasers is complex as optical components e.g. polarization
controllers (PC), laser gain medium, optical couplers, wavelength filters as well
as multiple cavity round trips, contribute to polarization evolution apart from
birefringence and dispersion of optical fibers. Ref. \cite{Zhao_10053} has
shown that the soliton polarization rotation is periodic where the period can
be multiple of cavity round trip time in a fiber ring laser. It has been shown
that soliton dynamics in a mode-locked fiber laser can produce either multiple
fixed polarization states occurring periodically or no fixed polarization
state at a fixed location of laser cavity \cite{Wu_046605}. This can lead to
different DOPs depending on the soliton dynamics. F8L is promising as this
fiber laser configuration can give rise to sub-picosecond laser pulses in ML
region \cite{Wang_265,Peng_731}. However, polarization instability
is a challenging issue in mode-locked F8L not based on
polarization-maintaining fibers.

In our experiment we employ the rotating quarter-wave plate method
\cite{goldstein2017polarized} to measure the DOP in both continuous wave (CW)
and ML regions. We find that for both cases the DOP can be tuned.

\section{Experimental setup}

\begin{figure}[ptb]
\begin{center}
\includegraphics[
		width=3.2396in, keepaspectratio
		]{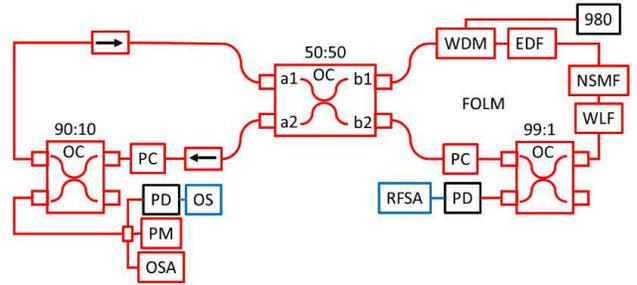}
\end{center}
\caption{Figure-8 laser. The FOLM on the right, and
the unidirectional loop on the left are connected by a 50:50 OC.}%
\label{FigSetup}%
\end{figure}

A sketch of the experimental setup is shown in Fig. \ref{FigSetup}. The F8L
consists of two fiber optical loops, which are connected through a $50:50$
optical coupler (OC). PCs are integrated into both loops.

The right loop serves as a fiber optical loop mirror (FOLM)
\cite{Doran_56,Mortimore_1217,Ibarra_191}. An erbium doped fiber (EDF) of
length $3.6%
%TCIMACRO{\unit{m}}%
%BeginExpansion
\operatorname{m}%
%EndExpansion
$ is integrated into the FOLM to provide gain in the telecom band. The EDF is pumped using a $980%
%TCIMACRO{\unit{nm}}%
%BeginExpansion
\operatorname{nm}%
%EndExpansion
$ diode laser and a wavelength division multiplexer (WDM). The
EDF is connected to a long nonlinear single mode fiber (NSMF) of length $50%
%TCIMACRO{\unit{m}}%
%BeginExpansion
\operatorname{m}%
%EndExpansion
$. Band selection is performed using a tunable wavelength filter (WLF) of
$0.42%
%TCIMACRO{\unit{nm}}%
%BeginExpansion
\operatorname{nm}%
%EndExpansion
$ width and $-3$ dB insertion loss. A $99:1$ OC in the right loop is connected
to a photodetector (PD), which is probed by a radio frequency spectrum
analyzer (RFSA).

For clockwise unidirectional propagation of light in the left loop, two
optical isolators (OI) are employed (labeled by arrows in Fig. \ref{FigSetup}%
). A 90:10 OC in the left loop is connected to an optical spectrum analyzer
(OSA), a PD, an oscilloscope (OS), and a rotating quarter-wave plate
polarimeter (PM)(Thorlabs PA430).

\begin{figure*}[ptb]
\begin{center}
\includegraphics[width=6.4in,keepaspectratio]{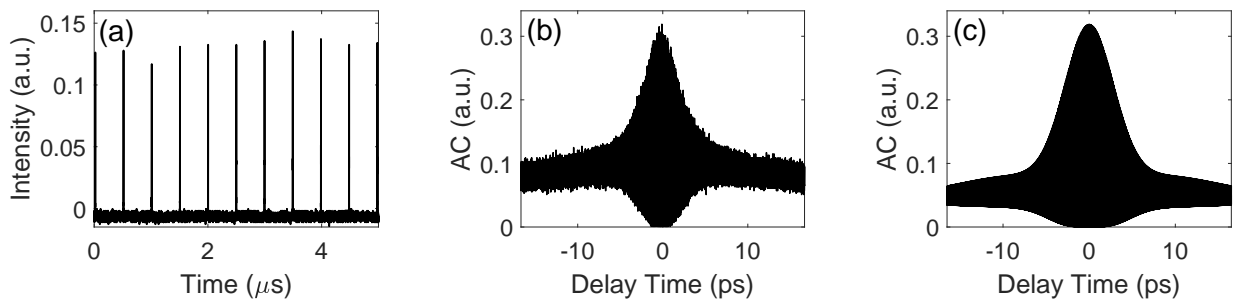}
\end{center}
\caption{Mode locking. (a) OS signal as a function of time at diode current
$I_{\mathrm{D}}=1\operatorname{A}$. (b) Measured AC as a function of delay
time at diode current $I_{\mathrm{D}}=1\operatorname{A}$. (c) Calculated AC
signal using Eq. (\ref{<IC>}) with pulse width of $\sigma_{\mathrm{c}%
}=1.8\operatorname{ps}$ and varying splitting time, which follows a normal
distribution having a standard deviation of $\sigma_{\mathrm{s}}%
=12\operatorname{ps}$.}%
\label{Fig_mode-locking}%
\end{figure*}

\section{Measurements}

The temporal F8L output signal measured by the OS in the region of ML is shown
in Fig. \ref{Fig_mode-locking}(a). Figure \ref{Fig_mode-locking}(b) shows the
corresponding autocorrelation (AC) signal as a function of a delay time.
Figure \ref{Fig_mode-locking}(c) shows the AC signal calculated using Eq.
(\ref{<IC>}) of appendix \ref{AppAC}. The calculated AC signal approximately
imitates the coherence artifact (spike) and broad pedestal (wings) of the
measured AC. The pulse width is extracted by fitting the data with Eq.
(\ref{<IC>}) (see caption of Fig. \ref{Fig_mode-locking}).

\begin{figure}[ptb]
\begin{center}
\includegraphics[width=3.2in,keepaspectratio]{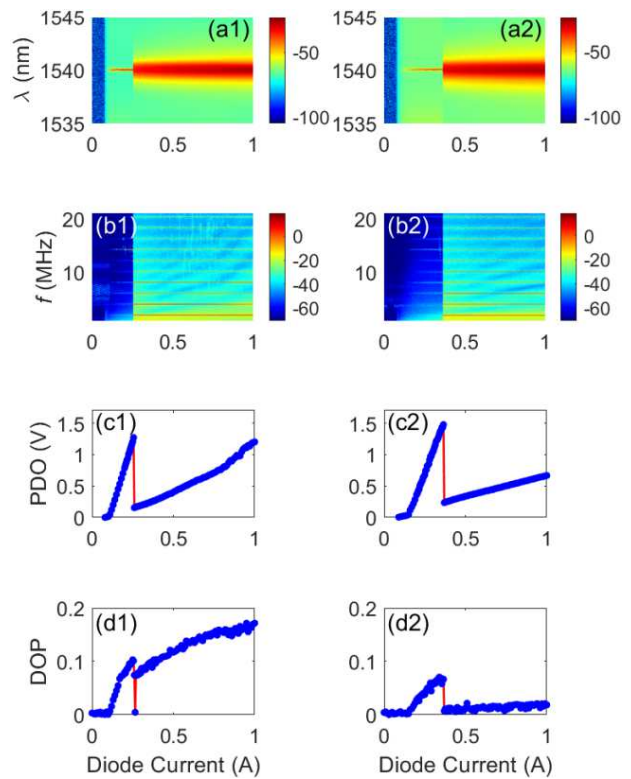}
\end{center}
\caption{Dependency on diode current $I_{\mathrm{D}}$. (a) Optical spectrum (in dBm units) at wavelength $\lambda$, (b) RF spectrum
(in dBm units) at frequency $\emph{f}$, (c) photodetector output power (PDO) and
(d) DOP. Left (right) panels, which are labeled by '1' ('2'), corresponds to maximized (minimized) DOP in ML regime. Note that the gain of
the PD employed to measure the RF spectrum is set at $10^{6}$ with bandwidth
of $3 \operatorname{MHz}$. The PD limited bandwidth gives rise to low pass
filtering artifact in the data presented in (b1) and (b2).}%
\label{Fig_OSA}%
\end{figure}

Plots of the optical spectrum measured by the OSA, the RF spectrum measured by
the RFSA, the averaged optical power measured by the PD in the right loop, and
the DOP measured by the PM, are shown in Fig. \ref{Fig_OSA} as a function of
the diode current $I_{\mathrm{D}}$. For the measurements shown in Fig.
\ref{Fig_OSA}(a1), (b1), (c1) and (d1), both PCs are tuned to maximize the DOP
in the ML regime, whereas the DOP in the ML regime is minimized for the
measurements shown in Fig. \ref{Fig_OSA}(a2), (b2), (c2) and (d2). For the
first (second) case the CW threshold occurs at diode current of $I_{\mathrm{D}%
}=0.1%
%TCIMACRO{\unit{A}}%
%BeginExpansion
\operatorname{A}%
%EndExpansion
$ ($I_{\mathrm{D}}=0.15%
%TCIMACRO{\unit{A}}%
%BeginExpansion
\operatorname{A}%
%EndExpansion
$), and the ML threshold at $I_{\mathrm{D}}=0.26%
%TCIMACRO{\unit{A}}%
%BeginExpansion
\operatorname{A}%
%EndExpansion
$ ($I_{\mathrm{D}}=0.37%
%TCIMACRO{\unit{A}}%
%BeginExpansion
\operatorname{A}%
%EndExpansion
$). As can be seen by comparing Fig. \ref{Fig_OSA}(d1) and (d2), DOP in the ML
regime can be tuned. The DOP plot in Fig. \ref{Fig_OSA}(d1) reveals a sharp
change in the state of polarization (SOP) at the transition from CW to ML
regime occurring at diode current of $I_{\mathrm{D}}=0.26%
%TCIMACRO{\unit{A}}%
%BeginExpansion
\operatorname{A}%
%EndExpansion
$. For this case, the DOP increases with diode current $I_{\mathrm{D}}$ in
both CW and ML regions. A SOP change occurring at the transition from CW to ML
regions is also seen in Fig. \ref{Fig_OSA}(d2). However, for this case, the
DOP remains very low above the transition.

\section{FOLM}

To account for the experimental observations, scattering and birefringence in
the FOLM are theoretically studied. The 50:50 OC is characterized by forward
(backward) transmission $t$ ($t^{\prime}$) and reflection $r$ ($r^{\prime}$)
amplitudes. Time reversal and mirror symmetries together with unitarity imply
that $t^{\prime}=t$, $r^{\prime}=r=it\left\vert r/t\right\vert $, and
$\left\vert t\right\vert ^{2}+\left\vert r\right\vert ^{2}=1$.

Consider light having SOP $\left\vert p_{\mathrm{i}}\right\rangle $ injected
into port a1 of the 50:50 OC. The transmitted (reflected) state $\left\vert
p_{\mathrm{T}}\right\rangle $ ($\left\vert p_{\mathrm{R}}\right\rangle $) at
port a2 (a1) is given by \cite{Buks_014421} $\left\vert p_{\mathrm{T}%
}\right\rangle =g\mathcal{A}_{\mathrm{T}}\left\vert p_{\mathrm{i}%
}\right\rangle $ ($\left\vert p_{\mathrm{R}}\right\rangle =g\mathcal{A}%
_{\mathrm{R}}\left\vert p_{\mathrm{i}}\right\rangle $), where the total gain
$g$ is real, the operator $\mathcal{A}_{\mathrm{T}}$ ($\mathcal{A}%
_{\mathrm{R}}$) is given by $\mathcal{A}_{\mathrm{T}}=tt^{\prime}e^{i\Theta
/2}\sigma_{z}J_{+}+rr^{\prime}e^{-i\Theta/2}J_{-}\sigma_{z}$ ($\mathcal{A}%
_{\mathrm{R}}=tr^{\prime}e^{i\Theta/2}\sigma_{z}J_{+}+rt^{\prime}%
e^{-i\Theta/2}J_{-}\sigma_{z}$), the nonlinear phase shift $\Theta$ is real,
$\sigma_{z}=\text{diag}\left(  1,-1\right)  $ is the Pauli's $z$ matrix, and
$J_{+}$ ($J_{-}$) is the Jones matrix corresponding to circulating the FOLM
loop in the clockwise (counter clockwise) direction. The assumption that both
$J_{+}$ and $J_{-}$ are unitary, i.e. $J_{+}^{\dag}J_{+}=J_{-}^{\dag}J_{-}=1$,
together with the relations%
\begin{align}
\mathcal{A}_{\mathrm{T}}^{\dag}\mathcal{A}_{\mathrm{T}}  &  =\mathcal{J}%
^{\dag}\left(
\begin{array}
[c]{c}%
t^{\ast}t^{\prime\ast}\\
r^{\ast}r^{\prime\ast}%
\end{array}
\right)  \left(
\begin{array}
[c]{cc}%
tt^{\prime} & rr^{\prime}%
\end{array}
\right)  \mathcal{J}\mathbf{\;,}\nonumber\\
&  =\mathcal{J}^{\dag}\left(
\begin{array}
[c]{c}%
\left\vert t\right\vert ^{2}\\
-\left\vert r\right\vert ^{2}%
\end{array}
\right)  \left(
\begin{array}
[c]{cc}%
\left\vert t\right\vert ^{2} & -\left\vert r\right\vert ^{2}%
\end{array}
\right)  \mathcal{J}\mathbf{\;,}\nonumber\\
&
\end{align}
and%
\begin{align}
\mathcal{A}_{\mathrm{R}}^{\dag}\mathcal{A}_{\mathrm{R}}  &  =\mathcal{J}%
^{\dag}\left(
\begin{array}
[c]{c}%
t^{\ast}r^{\prime\ast}\\
r^{\ast}t^{\prime\ast}%
\end{array}
\right)  \left(
\begin{array}
[c]{cc}%
tr^{\prime} & rt^{\prime}%
\end{array}
\right)  \mathcal{J}\nonumber\\
&  =\left\vert rt\right\vert ^{2}\mathcal{J}^{\dag}\left(
\begin{array}
[c]{c}%
1\\
1
\end{array}
\right)  \left(
\begin{array}
[c]{cc}%
1 & 1
\end{array}
\right)  \mathcal{J}\mathbf{\;,}\nonumber\\
&
\end{align}
where
\begin{equation}
\mathcal{J}^{\dag}=\left(  e^{-\frac{i\Theta}{2}}J_{+}^{\dag}\sigma
_{z},e^{\frac{i\Theta}{2}}\sigma_{z}J_{-}^{\dag}\right)  \mathbf{\;,}%
\end{equation}
yield
\begin{align}
\mathcal{A}_{\mathrm{T}}^{\dag}\mathcal{A}_{\mathrm{T}}  &  =\left(
\left\vert t\right\vert ^{2}-\left\vert r\right\vert ^{2}\right)
^{2}+2\left\vert rt\right\vert ^{2}\left(  1-S\right)  \mathbf{\;,}\\
\mathcal{A}_{\mathrm{R}}^{\dag}\mathcal{A}_{\mathrm{R}}  &  =2\left\vert
rt\right\vert ^{2}\left(  1+S\right)  \mathbf{\;,}%
\end{align}
where the operator $S$ is given by (note that $\sigma_{z}^{2}=1$)%
\begin{equation}
S=\frac{e^{-i\Theta}J_{+}^{\dag}\sigma_{z}J_{-}\sigma_{z}+e^{i\Theta}%
\sigma_{z}J_{-}^{\dag}\sigma_{z}J_{+}}{2}\mathbf{\;.}%
\end{equation}
These relations imply the unitarity condition
\begin{equation}
\mathcal{A}_{\mathrm{T}}^{\dag}\mathcal{A}_{\mathrm{T}}+\mathcal{A}%
_{\mathrm{R}}^{\dag}\mathcal{A}_{\mathrm{R}}=\left(  \left\vert t\right\vert
^{2}-\left\vert r\right\vert ^{2}\right)  ^{2}+4\left\vert rt\right\vert
^{2}=1\mathbf{\;.}%
\end{equation}

A derivation (see appendix \ref{AppIE}) similar to the one that yields the
Heisenberg uncertainty principle is employed to derive a lower bound upon the
FOLM reflectivity $P_{\mathrm{R}}=\left\langle p\right\vert \mathcal{A}%
_{\mathrm{R}}^{\dag}\mathcal{A}_{\mathrm{R}}\left\vert p\right\rangle
/\left\langle p\right.  \left\vert p\right\rangle $\ (commutation relation is
denoted by $\left[  \cdot,\cdot\right]  _{-}$)%
\begin{equation}
P_{\mathrm{R}}\geq\frac{\left\vert rt\right\vert ^{2}\left\vert \text{Re}%
\left\langle p\right\vert e^{-i\Theta}\left[  J_{+}^{\dag}\sigma_{z}%
,J_{-}\sigma_{z}\right]  _{-}\left\vert p\right\rangle \right\vert
}{\left\langle p\right.  \left\vert p\right\rangle }\;. \label{AR UB}%
\end{equation}
The relation $\mathcal{A}_{\mathrm{T}}^{\dag}\mathcal{A}_{\mathrm{T}%
}+\mathcal{A}_{\mathrm{R}}^{\dag}\mathcal{A}_{\mathrm{R}}=1$ together with
inequality (\ref{AR UB}) yields an upper bound upon the FOLM transmissivity
$P_{\mathrm{T}}=\left\langle p\right\vert \mathcal{A}_{\mathrm{T}}^{\dag
}\mathcal{A}_{\mathrm{T}}\left\vert p\right\rangle /\left\langle p\right.
\left\vert p\right\rangle $. \ However, both $P_{\mathrm{R}}$ and
$P_{\mathrm{T}}$\ become unbounded when $\left[  J_{+}^{\dag}\sigma_{z}%
,J_{-}\sigma_{z}\right]  _{-}=0$. In particular, no bound is imposed by the
inequality (\ref{AR UB}) when $J_{-}=\sigma_{z}J_{+}\sigma_{z}$. As will be
shown below, the condition $J_{-}=\sigma_{z}J_{+}\sigma_{z}$ is satisfied when
the SOP evolution along the FOLM depends only on the geometry of the fiber
spacial curve.

\section{SOP evolution}

The Jones matrices $J_{+}$ and $J_{-}$ can be calculated by integrating the
equation of motion for the SOP along the FOLM. Consider an optical fiber
winded in some spacial curve in space. Let $\mathbf{r}\left(  s\right)  $ be
an arc-length parametrization of this curve, i.e. the tangent $\mathbf{\hat
{s}=}\mathrm{d}\mathbf{r}/\mathrm{d}s$ is a unit vector. The normal and
binormal Serret - Frenet unit vectors are denoted by $\boldsymbol{\hat{\nu}}$
and $\mathbf{\hat{b}}$, respectively. The curve torsion $\tau$ is defined as
$\mathrm{d}\mathbf{\hat{b}/}\mathrm{d}s=-\tau\boldsymbol{\hat{\nu}}$
\cite{Buks_628}. A unit vector parallel to the electric field is denoted by
$\mathbf{\hat{e}}_{0}$. In the Serret - Frenet frame the unit vector
$\mathbf{\hat{e}}_{0}$, which is expressed as $\mathbf{\hat{e}}_{0}=e_{\nu
}\boldsymbol{\hat{\nu}}+e_{b}\mathbf{\hat{b}}$, evolves according to the
parallel transport equation of geometrical optics
\cite{Rytov_263,Pancharatnam_398,kravtsov1990geometrical}%
\begin{equation}
\frac{\mathrm{d}}{\mathrm{d}s}\left(
\begin{array}
[c]{c}%
e_{\nu}\\
e_{b}%
\end{array}
\right)  =i\mathcal{K}\left(
\begin{array}
[c]{c}%
e_{\nu}\\
e_{b}%
\end{array}
\right)  \mathbf{\;,} \label{e eom}%
\end{equation}
where $\mathcal{K}=\mathcal{K}_{\mathrm{g}}+\mathcal{K}_{\mathrm{f}}$. The
geometrical birefringence $\mathcal{K}_{g}$ is given by
\cite{Pancharatnam_398,Ross_455}%
\begin{equation}
\mathcal{K}_{\mathrm{g}}=\tau\left(
\begin{array}
[c]{cc}%
0 & -i\\
i & 0
\end{array}
\right)  \;,
\end{equation}
whereas $\mathcal{K}_{\mathrm{f}}$ is the fiber birefringence induced by
elasto-optic, electro-optic, or magneto-optic effects. For a lossless fiber
the matrix $\mathcal{K}$ is Hermitian.

Consider the case where $\mathcal{K}_{\mathrm{f}}$ vanishes. For this case the
solution to Eq. (\ref{e eom}) is given by \cite{Buks_628}
\begin{equation}
\left(
\begin{array}
[c]{c}%
e_{\nu}\left(  s\right) \\
e_{b}\left(  s\right)
\end{array}
\right)  =\left(
\begin{array}
[c]{cc}%
\cos\theta & \sin\theta\\
-\sin\theta & \cos\theta
\end{array}
\right)  \left(
\begin{array}
[c]{c}%
e_{\nu}\left(  0\right) \\
e_{b}\left(  0\right)
\end{array}
\right)  \mathbf{\;,}%
\end{equation}
where the rotation angle $\theta$ is given by the integrated torsion along the
fiber curve%
\begin{equation}
\theta=\int_{0}^{s}\mathrm{d}s^{\prime}\tau\left(  s^{\prime}\right)
\mathbf{\;}. \label{int torsion}%
\end{equation}
For the case where the curve end points are parallel, i.e. $\mathbf{\hat{s}%
}\left(  s\right)  =\mathbf{\hat{s}}\left(  0\right)  $, the following holds
$\theta=\Omega$, where $\Omega$ is the solid angle subtends by the closed
curve $\mathbf{\hat{s}}\left(  s^{\prime}\right)  $ at the origin. This
geometrical rotation of polarization, which is closely related to the Berry's
phase \cite{Berry_45}, has been experimentally measured in
\cite{Ross_455,tomita_937}.

For the case $\mathcal{K}_{\mathrm{f}}=0$ the following holds $J_{-}%
=\sigma_{z}J_{+}\sigma_{z}$, and consequently $S=\cos\Theta$, $\mathcal{A}%
_{\mathrm{T}}^{\dag}\mathcal{A}_{\mathrm{T}}=\left(  \left\vert t\right\vert
^{2}-\left\vert r\right\vert ^{2}\right)  ^{2}+4\left\vert rt\right\vert
^{2}\sin^{2}\left(  \Theta/2\right)  $ and $\mathcal{A}_{\mathrm{R}}^{\dag
}\mathcal{A}_{\mathrm{R}}=4\left\vert rt\right\vert ^{2}\cos^{2}\left(
\Theta/2\right)  $ (it is assumed that $J_{+}^{\dag}J_{+}=1$). Hence, for this
case both reflectivity $P_{\mathrm{R}}$ and transmissivity $P_{\mathrm{T}}$
becomes independent on the input SOP. For a 3 dB OC, i.e. $\left\vert
t\right\vert ^{2}=\left\vert r\right\vert ^{2}=1/2$, and in the linear limit,
i.e. for $\Theta=0$, the FOLM for the same case where $\mathcal{K}%
_{\mathrm{f}}=0$ becomes perfectly reflecting, i,e, $P_{\mathrm{T}}=0$ and
$P_{\mathrm{R}}=1$.

For treating the general case (where $\mathcal{K}_{\mathrm{f}}$ cannot be
disregarded, and consequently FOLM perfect reflectivity can not be guaranteed)
it is convenient to express the matrix $J_{-}$ as $J_{-}=J_{\mathrm{m}}%
\sigma_{z}J_{+}\sigma_{z}$, where $J_{\mathrm{m}}$ is unitary, i.e.
$J_{\mathrm{m}}^{\dag}J_{\mathrm{m}}=1$. Using this notation one finds that%
\begin{align}
P_{\mathrm{R}}  &  =4\left\vert rt\right\vert ^{2}\left(  1-\eta\right)
\mathbf{\;},\\
P_{\mathrm{T}}  &  =\left(  \left\vert t\right\vert ^{2}-\left\vert
r\right\vert ^{2}\right)  ^{2}+4\left\vert rt\right\vert ^{2}\eta\mathbf{\;},
\end{align}
where%
\begin{equation}
\eta=\sin^{2}\frac{\Theta}{2}-\frac{\text{Re}\left(  e^{-i\Theta}\left\langle
p\right\vert J_{+}^{\dag}V_{\mathrm{m}}J_{+}\left\vert p\right\rangle \right)
}{2\left\langle p\right.  \left\vert p\right\rangle }\mathbf{\;},
\end{equation}
and where $V_{\mathrm{m}}=\sigma_{z}\left(  J_{\mathrm{m}}-1\right)
\sigma_{z}$. Note that $J_{\mathrm{m}}=1$ and $V_{\mathrm{m}}=0$\ for the case
$\mathcal{K}_{\mathrm{f}}=0$.

\section{Mapping}

In this section a cycle to cycle mapping is derived to analyze the SOP
evolution. Stability analysis of the mapping allows the DOP evaluation in
steady state. The input SOP $\left\vert p_{\mathrm{i}}\right\rangle $ after
$n$ cycles is denoted by $\left\vert p_{n}\right\rangle $. The mapping between
$\left\vert p_{n}\right\rangle $ and $\left\vert p_{n+1}\right\rangle $ is
given by%
\begin{equation}
\left\vert p_{n+1}\right\rangle =gJ_{\mathrm{L}}\mathcal{A}_{\mathrm{T}%
}\left\vert p_{n}\right\rangle \mathbf{\;}, \label{p_n+1}%
\end{equation}
where the Jones matrix of the clockwise unidirectional left loop, which is
denoted by $J_{\mathrm{L}}$, is assumed to be unitary, i.e. $J_{\mathrm{L}%
}^{\dag}J_{\mathrm{L}}=1$. The mapping (\ref{p_n+1}) is nonlinear, since
$\mathcal{A}_{\mathrm{T}}$ depends on the nonlinear phase $\Theta$, which, in
turn, depends on the intensity $I_{n}=\left\langle p_{n}\right.  \left\vert
p_{n}\right\rangle $. In addition, the gain $g$ may vary due to saturation. In
general, the nonlinear phase $\Theta$ may also depend on the SOP, however,
this dependency is expected to be relatively weak since our setup is based on
regular (rather than polarization maintaining) single mode fibers. The
amplification factor $I_{n+1}/I_{n}$ corresponding to the SOP $\left\vert
p_{n}\right\rangle $ is denoted by $\mathcal{F}\left(  \left\vert
p_{n}\right\rangle \right)  $, where $\mathcal{F}\left(  \left\vert
p\right\rangle \right)  =g^{2}\left\langle p\right\vert \mathcal{A}%
_{\mathrm{T}}^{\dag}\mathcal{A}_{\mathrm{T}}\left\vert p\right\rangle
/\left\langle p\right.  \left\vert p\right\rangle $.

\begin{figure}[ptb]
\begin{center}
\includegraphics[
		width=3.2396in, keepaspectratio
		]{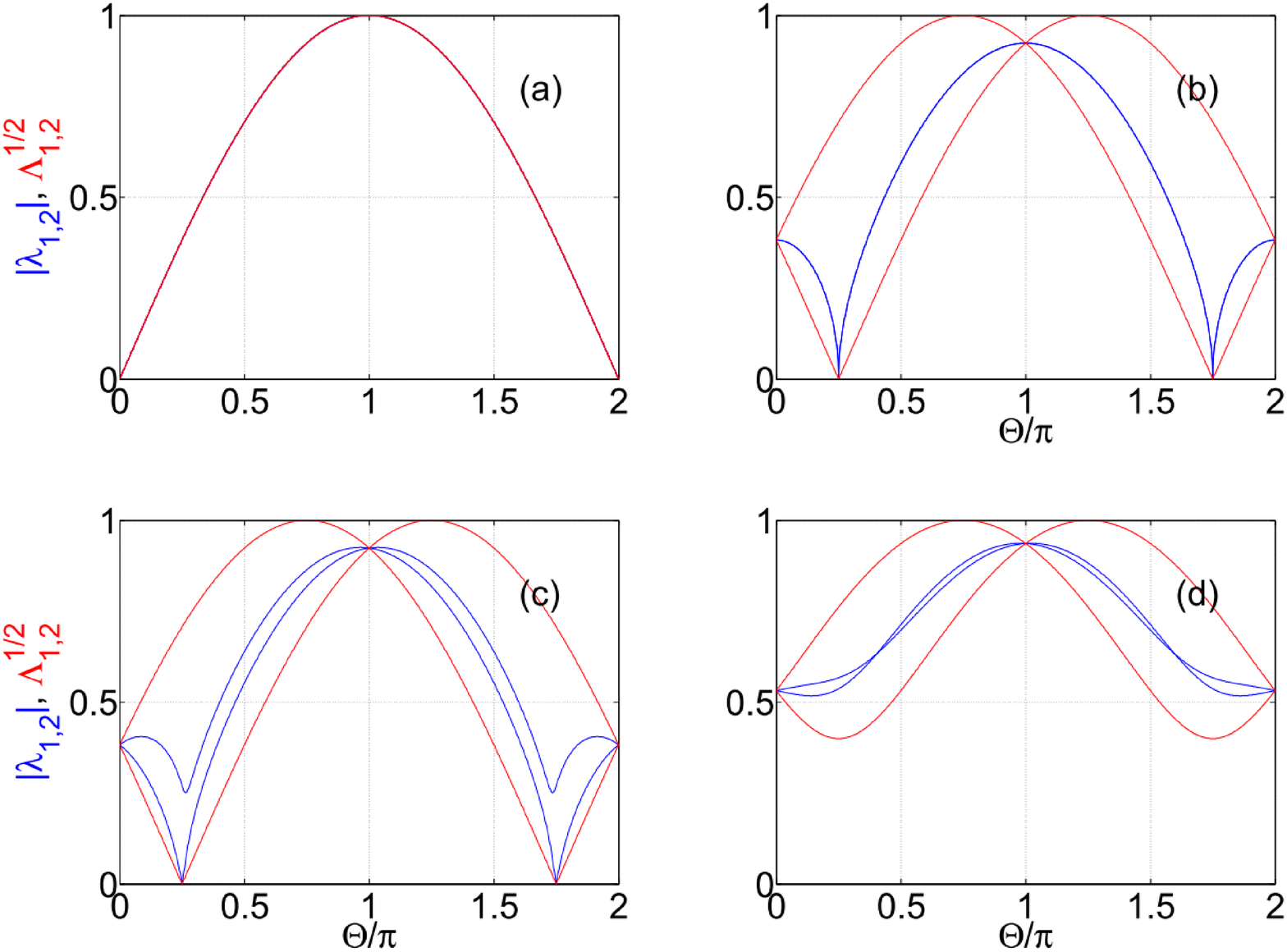}
\end{center}
\caption{Eigenvalues of $J_{\mathrm{L}}\mathcal{A}_{\mathrm{T}}$ (blue) and
$\mathcal{A}_{\mathrm{T}}^{\dag}\mathcal{A}_{\mathrm{T}}$ (red) as a function
of the nonlinear phase $\Theta$. (a) The following values are assumed for the
calculation $\theta_{+}=0$, $\varphi_{+}=0$ , $\phi_{+}=0$, $\theta
_{\mathrm{m}}=0.5\pi$, $\varphi_{\mathrm{m}}=0$, $\phi_{\mathrm{m}}=0$,
$\theta_{\mathrm{L}}=0$, $\varphi_{\mathrm{L}}=0$, $\phi_{ \mathrm{L}}=0$, and
$t=\left(  1/2\right)  ^{1/2}$. In the description below for the other panels
only modified (with respect to previous panels) parameters are specified. (b)
$\phi_{\mathrm{m}}=-0.5\pi$. (c) $\theta_{\mathrm{L}}=0.25\pi$, $\varphi_{
\mathrm{L}}=0.5\pi$ and $\phi_{\mathrm{L}}=0.25\pi$. (d) $t=\left(
0.3\right)  ^{1/2}$.}%
\label{Figev}%
\end{figure}

\begin{figure}[ptb]
\begin{center}
\includegraphics[
		width=3.2396in, keepaspectratio
		]{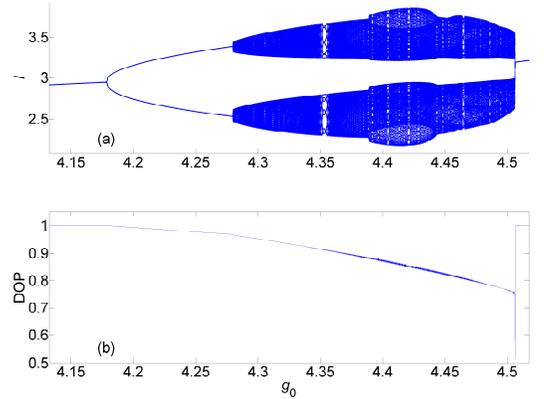}
\end{center}
\caption{A bifurcations diagram. The intensity $I$ (a) and DOP (b)
are plotted as a function of the small signal gain $g_{0}$. The
parameters assumed for the calculations are $\phi_{+}=0$, $\theta_{\mathrm{m}%
}=0.5 \pi$, $\varphi_{\mathrm{m}}=0$, $\phi_{\mathrm{m} }=-0.5\pi$,
$\theta_{\mathrm{L}}=0.5\pi$, $\varphi_{\mathrm{L}}=0.5\pi$, $\phi_{\mathrm{L}
}=0.5\pi$, $t=\left(  1/2\right)  ^{1/2}$, $\zeta=1$\ and $I_{\mathrm{s}}=1$.}%
\label{FigIvsg0}%
\end{figure}

The blue lines in Fig. \ref{Figev} represent the values of $\left\vert
\lambda_{1}\right\vert $ and $\left\vert \lambda_{2}\right\vert $ as a
function of $\Theta$ for four different cases, where $\lambda_{1,2}$ are the
eigenvalues of $J_{\mathrm{L}}\mathcal{A}_{\mathrm{T}}$. The unitary matrices
$J_{+}$, $J_{\mathrm{m}}$\ and $J_{\mathrm{L}}$ are expressed as
$J_{+}=U\left(  \theta_{+},\varphi_{+},\phi_{+}\right)  $, $J_{\mathrm{m}%
}=U\left(  \theta_{\mathrm{m}},\varphi_{\mathrm{m}},\phi_{\mathrm{m}}\right)
$ and $J_{\mathrm{L}}=U\left(  \theta_{\mathrm{L}},\varphi_{\mathrm{L}}%
,\phi_{\mathrm{L}}\right)  $, respectively, where%
\begin{align}
&  U\left(  \theta,\varphi,\phi\right) \nonumber\\
&  =\left(
\begin{array}
[c]{cc}%
\cos\frac{\phi}{2}-i\cos\theta\sin\frac{\phi}{2} & -i\sin\theta e^{-i\varphi
}\sin\frac{\phi}{2}\\
-i\sin\theta e^{i\varphi}\sin\frac{\phi}{2} & \cos\frac{\phi}{2}+i\cos
\theta\sin\frac{\phi}{2}%
\end{array}
\right)  \mathbf{\;},\nonumber\\
&
\end{align}
and where all angles $\theta$, $\varphi$ and $\phi$ are real. The matrix
$U\left(  \theta,\varphi,\phi\right)  $ represents SOP rotation around the
unit vector $\left(  \sin\theta\cos\varphi,\sin\theta\sin\varphi,\cos
\theta\right)  $ by an angle $\phi$. The red lines in Fig. \ref{Figev}
represent the values of $\Lambda_{1}^{1/2}$ and $\Lambda_{2}^{1/2}$, where
$\Lambda_{1,2}$ are the non-negative eigenvalues of the Hermitian and
positive-definite operator $\mathcal{A}_{\mathrm{T}}^{\dag}\mathcal{A}%
_{\mathrm{T}}$. Let $\left\vert p_{\mathrm{F}}\right\rangle $ be a SOP, for
which the amplification factor $\mathcal{F}\left(  \left\vert p\right\rangle
\right)  $ is maximized (for a given $\Theta$). If, in addition, $\left\vert
p_{\mathrm{F}}\right\rangle $ is an eigenvector of the mapping operator
$gJ_{\mathrm{L}}\mathcal{A}_{\mathrm{T}}$, then the DOP is expected to be
relatively high in steady state, provided that $\left\vert p_{\mathrm{F}%
}\right\rangle $ is a locally stable fixed point of the mapping (\ref{p_n+1}).
For the case shown in Fig. \ref{Figev}(a), for which the operator
$\mathcal{A}_{\mathrm{T}}$ is given by$\mathcal{A}_{\mathrm{T}}=i\sin\left(
\Theta/2\right)  \sigma_{z}$, the SOP $\left\vert p_{\mathrm{F}}\right\rangle
$ is an eigenvector of $gJ_{\mathrm{L}}\mathcal{A}_{\mathrm{T}}$ for all
$\Theta$ [note that $\left\vert \lambda_{1,2}\right\vert =\Lambda_{1,2}%
^{1/2}=\sin\left(  \Theta/2\right)  $ for this case]. The equality $\left\vert
\lambda_{1,2}\right\vert =\Lambda_{1,2}^{1/2}$ [see Fig. \ref{Figev}(a)] is
violated by the transformation $J_{\mathrm{m}}$ [see Fig. \ref{Figev}(b)], and
the equality $\left\vert \lambda_{1}\right\vert =\left\vert \lambda
_{2}\right\vert $ [see Fig. \ref{Figev}(a) and (b)] is violated by the
transformation $J_{\mathrm{L}}$ [see Fig. \ref{Figev}(c)]. The effect of
violating the condition $\left\vert t\right\vert ^{2}=\left\vert r\right\vert
^{2}$ is demonstrated in Fig. \ref{Figev}(d).

The drop in PDO [see Fig. \ref{Fig_OSA}(c1) and (c2)] and DOP [see
Fig. \ref{Fig_OSA}(d1) and (d2)] occurring in the transition from CW to ML is
attributed to the change in peak power, which in turn, affects SOP evolution
due to nonlinearity of the cycle to cycle mapping \ref{p_n+1} (induced by both Kerr effect and gain
saturation). The drop in PDO
represents the change in the averaged optical loss per cycle, which depends on
the FOLM reflectivity $P_{\mathrm{R}}$.
A bifurcation diagram example of the mapping (\ref{p_n+1}) is shown in Fig.
\ref{FigIvsg0}. The parameters that are used for the calculation are listed in
the figure caption. To account for saturation, the gain $g$ is assumed to be
given by%
\begin{equation}
g=\frac{g_{0}}{1+\frac{I}{I_{\mathrm{s}}}}\mathbf{\;}, \label{g}%
\end{equation}
where $g_{0}$ is the small signal gain, and $I_{\mathrm{s}}$ is the saturation
intensity. The Kerr effect induced nonlinear phase $\Theta$ is assumed to be
given by $\Theta=\zeta I$, where $\zeta$ is a positive constant, and where
$I=\left\langle p\right.  \left\vert p\right\rangle $ is the intensity. For
the example shown in Fig. \ref{FigIvsg0}, a period doubling bifurcation occurs
at $g_{0}=4.18$. Relatively low DOP is expected in the region $4.28\leq
g_{0}\leq4.51$, where the period time of the mapping (\ref{p_n+1}) becomes
much longer than the loop fundamental period.

Note that the mapping \ref{p_n+1} strongly depends on the unitary
matrices $J_{+}$, $J_{\mathrm{m}}$\ and $J_{\mathrm{L}}$. In the experiment,
this dependency can be explored by tuning both PCs. The dependency shown in
Fig. \ref{Fig_OSA}, of both PDO and DOP on diode current $I_{\mathrm{D}}$, can
be significantly modified by retuning both PCs. Similarly, the bifurcation
diagram shown in Fig. \ref{FigIvsg0} can be significantly modified by changing
the assumed unitary matrices. In a a fiber-based setup, it is very difficult
to independently measure each of these unitary matrices, for any given paddle
configuration of both PCs. However, when these unitary matrices are treated as
fitting parameters, good agreement between theory and experiment can be
obtained.

\section{Summary}

In summary, DOP tunability is experimentally demonstrated in a
F8L, in both CW and ML regions. A mapping (\ref{p_n+1}) is employed to derive
the SOP time evolution. High DOP can be obtained in the regions where the
mapping has a locally stable fixed point. Nonlinearity of the cycle to cycle
SOP mapping \ref{p_n+1} gives rise to complex dynamics. The current
experimental setup, which is based on the rotating quarter-wave plate method,
does not allow monitoring the relatively fast SOP cycle to cycle evolution,
since the fiber ring period time is much shorter than the plate rotation
period time. Future experiments, which will employ a faster polarimeter, will
be devoted to the cycle to cycle complex dynamics of this system.

We thank A. Becker and V. Smulkovsky for technical help. This work was
supported by the Israeli science foundation, the Israeli ministry of science,
and by the Technion security research foundation.

The data that support the findings of this study are available from the corresponding author 
upon reasonable request.

\appendix

\section{Autocorrelation}

\label{AppAC}

In a Michelson interferometer with a second harmonic signal, the detector
signal $I_{\mathrm{C}}\left(  \tau\right)  $ corresponding to second-order
interferometric correlation can be written as a function of a delay time
$\tau$ as [see Eq. (9.6) of Ref. \cite{diels2006ultrashort}]%
\begin{equation}
I_{\mathrm{C}}\left(  \tau,E\left(  t\right)  \right)  =\int_{-\infty}%
^{\infty}\mathrm{d}t\;\left\vert \left(  E\left(  t\right)  +E\left(
t-\tau\right)  \right)  ^{2}\right\vert ^{2}\mathbf{\;}, \label{IC}%
\end{equation}
where, $E\left(  t\right)  =e^{i\omega_{0}t}f_{\sigma_{\mathrm{c}}}%
^{1/2}\left(  t\right)  $ is the electric field of a Gaussian partial pulse at
time $t$ [see Eq. (2.39) of Ref. \cite{trebino2000frequency}], $\omega_{0}$ is
the carrier angular frequency, and the normal distribution function having a
standard deviation $\sigma$ and a vanishing expectation value is given by%
\begin{equation}
f_{\sigma}\left(  x\right)  =\frac{1}{\sigma\sqrt{2\pi}}\exp\left(
-\frac{x^{2}}{2\sigma^{2}}\right)  \mathbf{\;}.
\end{equation}
The pulse full width half maxima is $2\sqrt{2}\left(  \ln^{1/2}2\right)
\sigma_{\mathrm{c}}$.

A fitting based on Eq. (\ref{IC}) could not accurately replicate the broad
pedestal (wings) in our experimental AC data shown in Fig.
\ref{Fig_mode-locking}(b). Better agreement can be obtained from a
double-pulse model, which assumes pulse splitting with a varying separation
time $t_{\mathrm{s}}$ (see Fig. 4.8 of Ref. \cite{trebino2000frequency}). The
splitting gives rise to the so-called coherence artifact. In this model the
electric field $E_{\mathrm{s}}\left(  t,t_{\mathrm{s}}\right)  $ is assumed to
be given by
\begin{equation}
E_{\mathrm{s}}\left(  t,t_{\mathrm{s}}\right)  =e^{i\omega_{0}t}%
\frac{f_{\sigma_{\mathrm{c}}}^{1/2}\left(  t-\frac{t_{\mathrm{s}}}{2}\right)
+f_{\sigma_{\mathrm{c}}}^{1/2}\left(  t+\frac{t_{\mathrm{s}}}{2}\right)  }%
{2}\mathbf{\;}.
\end{equation}
The double-pulse assumption is supported by Ref. \cite{Bale_23137}, which
demonstrates transition from single-pulsing to double-pulsing regime with
increasing pump current beyond the single-pulsing instability threshold. In
our case, the time of separation $t_{\mathrm{s}}$ is considered to have a
normal distribution with standard deviation $\sigma_{\mathrm{s}}$. The
averaged AC signal, which is denoted by $\left\langle I_{\mathrm{C}}\left(
\tau\right)  \right\rangle $, is given by%
\begin{equation}
\left\langle I_{\mathrm{C}}\left(  \tau\right)  \right\rangle =\int_{-\infty
}^{\infty}\mathrm{d}t_{\mathrm{s}}\;f_{\sigma_{\mathrm{s}}}\left(
t_{\mathrm{s}}\right)  I_{\mathrm{C}}\left(  \tau,E_{\mathrm{s}}\left(
t,t_{\mathrm{s}}\right)  \right)  \mathbf{\;}. \label{<IC>}%
\end{equation}
The values of both standard deviations $\sigma_{\mathrm{c}}$ and
$\sigma_{\mathrm{s}}$ are extracted by data fitting (see caption of Fig.
\ref{Fig_mode-locking}).

\section{Inequality (\ref{AR UB})}

\label{AppIE}

The positive definite operator $\mathcal{A}_{\mathrm{R}}^{\dag}\mathcal{A}%
_{\mathrm{R}}$ can be expressed as
\begin{equation}
\mathcal{A}_{\mathrm{R}}^{\dag}\mathcal{A}_{\mathrm{R}}=\frac{\left[
\mathcal{A}_{\mathrm{R}}^{\dag},\mathcal{A}_{\mathrm{R}}\right]  +\left[
\mathcal{A}_{\mathrm{R}}^{\dag},\mathcal{A}_{\mathrm{R}}\right]  _{+}}{2}\;,
\end{equation}
where $\left[  \mathcal{A}_{\mathrm{R}}^{\dag},\mathcal{A}_{\mathrm{R}%
}\right]  _{-}=\mathcal{A}_{\mathrm{R}}^{\dag}\mathcal{A}_{\mathrm{R}%
}-\mathcal{A}_{\mathrm{R}}^{\dag}\mathcal{A}_{\mathrm{R}}$ is anti-Hermitian
and $\left[  \mathcal{A}_{\mathrm{R}}^{\dag},\mathcal{A}_{\mathrm{R}}\right]
_{+}=\mathcal{A}_{\mathrm{R}}^{\dag}\mathcal{A}_{\mathrm{R}}+\mathcal{A}%
_{\mathrm{R}}\mathcal{A}_{\mathrm{R}}^{\dag}$ is Hermitian, thus%
\begin{equation}
\left\vert \frac{\left\langle p\right\vert \mathcal{A}_{\mathrm{R}}^{\dag
}\mathcal{A}_{\mathrm{R}}\left\vert p\right\rangle }{\left\langle p\right.
\left\vert p\right\rangle }\right\vert ^{2}\geq\left\vert \frac{\left\langle
p\right\vert \left[  \mathcal{A}_{\mathrm{R}}^{\dag},\mathcal{A}_{\mathrm{R}%
}\right]  _{-}\left\vert p\right\rangle }{2\left\langle p\right.  \left\vert
p\right\rangle }\right\vert ^{2}\;.
\end{equation}
The relation $\mathcal{A}_{\mathrm{R}}=tr^{\prime}e^{i\Theta/2}\sigma_{z}%
J_{+}+rt^{\prime}e^{-i\Theta/2}J_{-}\sigma_{z}$ yields (recall that it is
assumed that $J_{+}^{\dag}J_{+}=J_{-}^{\dag}J_{-}=1$)%
\begin{align}
\frac{\left[  \mathcal{A}_{\mathrm{R}}^{\dag},\mathcal{A}_{\mathrm{R}}\right]
_{-}}{\left\vert rt\right\vert ^{2}}  &  =e^{-i\Theta}\left\langle
p\right\vert \left[  J_{+}^{\dag}\sigma_{z},J_{-}\sigma_{z}\right]
_{-}\left\vert p\right\rangle \nonumber\\
&  +e^{i\Theta}\left\langle p\right\vert \left[  \sigma_{z}J_{-}^{\dag}%
,\sigma_{z}J_{+}\right]  _{-}\left\vert p\right\rangle \;,\nonumber\\
&
\end{align}
hence%
\begin{equation}
\frac{\left\vert \frac{\left\langle p\right\vert \mathcal{A}_{\mathrm{R}%
}^{\dag}\mathcal{A}_{\mathrm{R}}\left\vert p\right\rangle }{\left\langle
p\right.  \left\vert p\right\rangle }\right\vert ^{2}}{\left\vert
rt\right\vert ^{4}}\geq\left\vert \frac{\operatorname{Re}\left(  e^{-i\Theta
}\left\langle p\right\vert \left[  J_{+}^{\dag}\sigma_{z},J_{-}\sigma
_{z}\right]  _{-}\left\vert p\right\rangle \right)  }{\left\langle p\right.
\left\vert p\right\rangle }\right\vert ^{2}\;,
\end{equation}
and thus the inequality (\ref{AR UB}) holds.

\bibliographystyle{ieeepes}
\bibliography{acompat,Eyal_Bib}

\end{document}